# An efficient procedure to predict the acoustophoresis of axisymmetric irregular particles above ultrasound transducer array

Tianquan Tang[a,b,*], Lixi Huang[a,b]

[a]*Department of Mechanical Engineering, The University of Hong Kong, Pokfulam, Hong Kong SAR, China*
[b]*Lab for Aerodynamics and Acoustics, HKU Zhejiang Institute of Research and Innovation, 1623 Dayuan Road, Lin An District, Hangzhou, China*

\* Corresponding email: tianquan@connect.hku.hk.

Acoustic radiation force and torque arising from wave scattering are able to translate and rotate matter without contact. However, the existing research mainly focused on manipulating simple symmetrical geometries, neglecting the significance of geometric features. For the non-spherical geometries, the shape of the object strongly affects its scattering properties, and thus the radiation force and torque as well as the acoustophoretic process. Here, we develop a semi-analytical framework to calculate the radiation force and torque exerted on the axisymmetric particles excited by a user-customized transducer array based on a conformal transformation approach, capturing the significance of the geometric features. The derivation framework is established under the computation coordinate system (CCS), whereas the particle is assumed to be static. For the dynamic processes, the rotation of particle is converted as the opposite rotation of transducer array, achieved by employing a rotation transformation to tune the incident driving field in the CCS. Later, the obtained radiation force and torque in the CCS should be transformed back to the observation coordinate system (OCS) for force and torque analysis. The radiation force and torque exerted on particles with different orientations are validated by comparing the full three-dimensional numerical solution in different phase distributions. It is found that the proposed method presents superior computational accuracy, high geometric adaptivity, and good robustness to various geometric features, while the computational efficiency is more than 100 times higher than that of the full numerical method. Furthermore, it is found that the dynamic trajectories of particles with different geometric features are completely different, indicating that the geometric features can be a potential degree of freedom to tune acoustophoretic process. The ability to predict the acoustophoretic process of non-spherical particles above a user-customized transducer array has improved our understanding of the effect of shape asymmetry, which can also be used to verify the effectiveness of acoustic tweezers in manipulating non-spherical objects.

# 1. Introduction

Acoustic waves exert acoustic radiation force and torque on objects because of the momentum transfer that arises from acoustic scattering effects of the wave-particle interaction [1][2][3][4]; these second-order force and torque, caused by inherent nonlinearities in the governing physics [5], have raised great interest in applications, including particle assembly [6][7][8], acoustophoretic printing [9][10], and acoustic holograms [11], since they can perform biocompatible, contact-free, and precise manipulation. Functionally, these contactless manipulations can be divided into two major categories: transportation and rotation of objects. The transportation-related processes are of critical importance in droplet coalescence [12], chemical analysis [13], and volumetric display [14]. Differently, the rotational manipulation of objects [15] can reveal hidden structural details, which are not visible in translational manipulation. Hence, it is an effective tool to interrogate morphological phenotype [16] and to operate microsurgery [17] for microorganisms.

Single-sided transducer array [18][19] is one of the most common and effective arrangements in containerless transportation [19][20] or contactless rotation [21] of levitated particles in the air. For Rayleigh particles, the levitated objects can be simply regarded as spherical particles. The acoustic radiation force on these particles can be evaluated according to the gradient of the Gorkov potential [22]. The scattering contribution from Rayleigh particles is negligible, and thus the Gor'kov potential merely depends on the external driving fields from transducers. With proper spatial arrangement and operating parameters (such as retrieval algorithm [23]) of the transducers, multiple Gor'kov potential wells or acoustic vortices can be created to manipulate particles.

Beyond the Rayleigh regime, the scattering contribution becomes significant. Neglecting the geometric asymmetry, a set of semi-analytical expressions have been derived for the radiation force and torque based on the partial wave expansion series [24][25][26]. An obvious limitation of the above studies is they all assumed that the manipulated object(s) are spherical, indicating that the radial distance from the mass center of the object to the locus of any point on the object surface is a constant. In this way, the boundary condition can be conveniently employed to decouple each mode in the expansion series. Thus the scattered wavefields are obtainable by solving a system of linear equations. It is worth emphasizing that, in reality, most manipulated objects have a certain degree of asymmetry in their morphology. This simplification of the spherical shape neglects the effect of asymmetry, which is an indispensable factor in evaluating the radiation force and torque [27][28], thereby the underlying acoustophoresis. In fact, exact solutions can be found for only a limited class of geometries where separation of variables is applicable. In other words, the problem must be able to formulate in a specific coordinate system in which the locus of points corresponding to one of the coordinates (typically, the radial coordinate) being a constant coincides with the scatterer surface. Consequently, the Helmholtz wave equation specified by the coordinate-independent boundary conditions is solvable. For irregular objects, an alternative to calculate the acoustic radiation force and torque is the use of numerical techniques [28][29], while it is limited by high computational cost. More importantly, it is impractical or cumbersome to analyze the dynamics of the objects, i.e., the acoustophoretic process, since we have to continuously renew the particle positions and orientations based on the estimated radiation force and torque.

A promising framework to semi-analytically express the radiation force and torque is the use of the conformal transformation approach to map the physical asymmetric geometry into a sphere in a new mapping coordinate system [30][31][32], in which the locus of all points corresponding to the new radial coordinate being a constant exactly coincides with the scatterer surface. Thus the boundary conditions are able to enforce easily, and the corresponding scattered fields can be solved [33]. After the scattering field is known, the acoustic radiation force and torque can be asymptotically obtained. Undoubtedly, the above framework should be a viable route to estimate the acoustic radiation force and torque on an axisymmetric particle. However, it should be emphasized that the derivations are established under the particle system, whose origin and $z$-axis is set to coincide with, respectively, the mass center and symmetric axis of the particle (i.e., the computation system illustrated in Fig. 1). During the acoustophoresis, the positions and orientations of the non-spherical particles are constantly changing under the effects of the radiation force and torque, meaning that the particle system is a moving coordinate system. In contrast, the transducer array (or the observation system shown in Fig. 1) remains static. Consider that the computational framework for the radiation force and torque based on the conformal technique is established under the premise that the particle is static. We need to reconsider the physical background: the particle is fixed while the transducer array or the incident driving wavefield is constantly moving. Mathematically, this case is equivalent to the incident driving wavefield at rest, whereas the particle moves. Clearly, to predict the acoustophoretic process of non-spherical particles, skillfully and constantly translation and rotation transformations are needed to transform the incident driving wave between the static system (i.e., the observation system) and the moving system (i.e., the computational system) [34].

Our present work aims to present a general semi-analytical solution for the acoustic radiation force and torque exerted on an arbitrarily axisymmetric particle caused by a user-customized transducer array. Firstly, the translation and rotation transformations [34] are needed to reshape the wave function of transducers from the observation coordinate system (OCS) to the particle system or the computational coordinate system (CCS). Then, the conformal transformation approach [31] is employed to capture the effect of geometric features, which transforms the non-spherical surface into a spherical one. The boundary conditions are enforced, and the Helmholtz wave equation is solved. The radiation force and torque can be asymptotically derived by integrating the acoustic potential field on a far-field control surface under the CCS. Similarly, the translation and rotation transformations are employed to map the radiation force and torque from the CCS to the OCS. Combined with the viscous drag force and torque [35][36], the acoutophoresis of non-spherical particles under a transducer array can be predicted. The remainder of this paper is structured as follows: In Section 2, the mathematical formulations are given to evaluate the radiation force and torque, thereby the acoustophoretic process. The formulations start from the OCS and extend to the CCS for the radiation force and torque, while back to the OCS to predict the acoustophoresis of non-spherical particles. In Section 3, the computational performance of the proposed method is examined through a set of full three-dimensional numerical simulations in terms of the radiation force and torque exerted on different non-spherical particles. Furthermore, the acoustophoretic processes of spherical and non-spherical particles are visualized, compared, and discussed. Finally, some conclusions are given in Section 4. The geometric data of the non-spherical particles and the numerical model used to verify the semi-analytical framework are provided in the Supplementary Material.

## 2. Theoretical model

### 2.1. Computation and observation coordinate systems

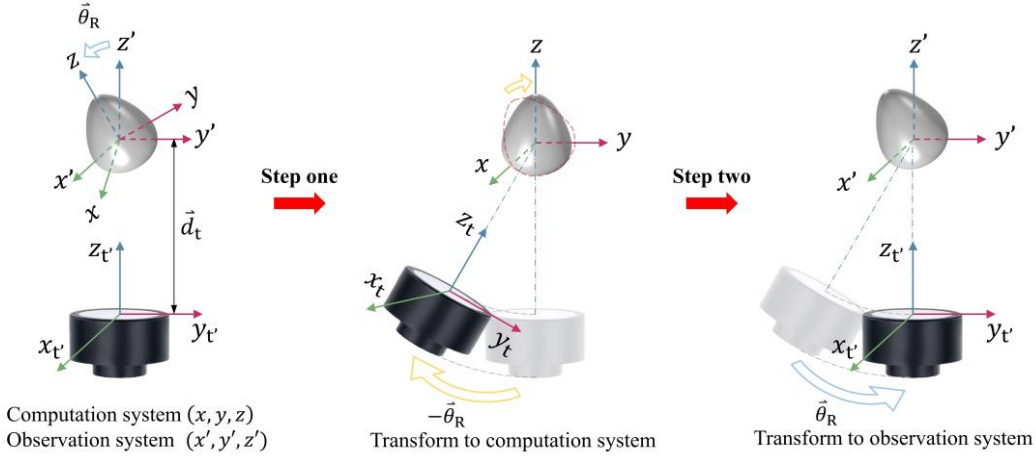

**Figure 1:** Rotation transformation between the computation coordinate system $(x, y, z)$ and the observation coordinate system $(x', y', z')$.

In a particle-transducer system, we define an observation coordinate system (OCS) where the origin coincides with the center of mass of the manipulated particle, denoted as $(x', y', z')$ system, which is acceptable as an absolute coordinate system to further discuss the dynamic problem since it is established under the well-known Cartesian coordinate system. By contrast, a computation coordinate system (CCS), denoted as $(x, y, z)$ system, is introduced to better characterize the axisymmetric particle. The origin of these two systems is spatially coincident, while the $z$-axis of the computation system is defined by the symmetric axis of the particle, as shown in Fig. 1. In this way, considering the axisymmetric physics, a general three-dimensional geometry can be equivalently described by a two-dimensional cross-sectional slice plane ($zOf$-plane) and an azimuthal coordinate variable ($\phi \in [0, 2\pi]$), as depicted later in Fig. 3. Since the boundary of any cross-sectional slice for any specified azimuthal angle is identical, the geometric features merely depend on the cross-sectional slice. This property enables the conformal transformation method to map a two-dimensional irregular cross-sectional slice in the CCS to a new quasi-spherical coordinate system [30], in where the locus of all points of the slice boundary is equal to a constant. Hence, the separation of variables can be used to solve the Helmholtz wave equation subjecting to the spherical boundary conditions in the new mapping coordinate system, and the acoustic radiation force and torque are obtainable analytically. Generally, the CCS does not map with the OCS as the non-spherical particles are continuously rotating, affected by the radiation torque. Figure 1 illustrates the rotational relationship between the CCS and the OCS. The CCS orientationally deviates from the OCS by a rotation angle $\vec{\theta}_R = (\theta_{x'}, \theta_{y'}, \theta_{z'})$, where $\theta_i, i = x', y', z'$ meaning the particle rotates along $i$-axis for a angle $\theta_i$, while its sign is determined by the right-hand rule. Mathematically, these two coordinate systems can be connected by applying corresponding rotation transformation matrix as

$$\mathbf{R}_x(\theta_{x'}) = \begin{bmatrix} 1 & 0 & 0 \\ 0 & \cos(\theta_{x'}) & -\sin(\theta_{x'}) \\ 0 & \sin(\theta_{x'}) & \cos(\theta_{x'}) \end{bmatrix}, \quad (1)$$

$$\mathbf{R}_y(\theta_{y'}) = \begin{bmatrix} \cos(\theta_{y'}) & 0 & \sin(\theta_{y'}) \\ 0 & 1 & 0 \\ -\sin(\theta_{y'}) & 0 & \cos(\theta_{y'}) \end{bmatrix},$$

$$\mathbf{R}_z(\theta_{z'}) = \begin{bmatrix} \cos(\theta_{z'}) & -\sin(\theta_{z'}) & 0 \\ \sin(\theta_{z'}) & \cos(\theta_{z'}) & 0 \\ 0 & 0 & 1 \end{bmatrix}.$$

Specifically, the coordinate variables between the CCS and the OCS can be mutually expressed using the rotation transformation matrix as

$$\begin{cases} [x, y, z] = [x', y', z'] \cdot \mathbf{R}_x(-\theta_{x'})\mathbf{R}_y(-\theta_{y'})\mathbf{R}_z(-\theta_{z'}) \\ [x', y', z'] = [x, y, z] \cdot \mathbf{R}_z^{-1}(\theta_{z'})\mathbf{R}_y^{-1}(\theta_{y'})\mathbf{R}_x^{-1}(\theta_{x'}) \end{cases}. \quad (2)$$

Based on Eq. (2), the coordinate variables and the derived radiation force and torque can be conveniently transformed between the CCS and the OCS.

## 2.2. Wave function of a single transducer

The circular piston radiator is an important example in ultrasonics as it is about the simplest approximation that can be made for radiation into an infinite medium from a circular ultrasound transducer [37][38]. We consider a time-harmonic far-field wave function of a circular piston source [38] with respect to observation transducer coordinate system, that is $(x_{t'}, y_{t'}, z_{t'})$ system illustrated in Fig. 1, as

$$\hat{p}(x_{t'}, y_{t'}, z_{t'}) = P_0 \cdot \frac{2j_1\left(\frac{kd}{2}\sin\left(\arccos\frac{z_{t'}}{R_{t'}}\right)\right)}{\frac{kd}{2}\sin\left(\arccos\frac{z_{t'}}{R_{t'}}\right)} \cdot \frac{e^{ikR_{t'}}}{R_{t'}}, \quad (3)$$

where position abbreviation $R_{t'} = \sqrt{x_{t'}^2 + y_{t'}^2 + z_{t'}^2}$ and the power parameter $P_0 = \frac{-i\rho_0 c_s k d^2 \hat{v}_0}{8}$. The hat symbol ˆ represents the complex amplitude of the corresponding variable. Parameters $\rho_0$ and $c_s$ are density and adiabatic speed of sound of a homogeneous host fluid, respectively. The wavenumber of fluid $k = \frac{\omega}{c_s}$ with angular frequency $\omega$. The transducer is characterized by its diameter $d$ and complex amplitude of the radial velocity $\hat{v}_0$. Function $j_1(\cdot)$ represents the Bessel function of the first kind.

For the problem under consideration, benefitting from the axisymmetric property of a particle, all the derivations are established under the CCS, which indicates that the wave function should be transformed and re-expressed using coordinate variables of the $(x, y, z)$ system. Under the external forces and torques, we assume that the particle has rotated at an angle of $\vec{\theta}_R$ and translated to the position of $\vec{r}_t = (r_{t,x'}, r_{t,y'}, r_{t,z'})$ relative to the OCS. A rotation transformation and a translation transformation are required to map the wave function of Eq. (3) to the CCS (corresponding to step one in Fig. 1), which yields

$$\begin{cases} \hat{p}(x, y, z) = P_0 \cdot \dfrac{2j_1\left(\frac{kd}{2}\sin\left(\arccos\frac{z_{t'}}{R_{t'}}\right)\right)}{\frac{kd}{2}\sin\left(\arccos\frac{z_{t'}}{R_{t'}}\right)} \cdot \dfrac{e^{ikR_{t'}}}{R_{t'}} \\ [x_{t'}, y_{t'}, z_{t'}] = [x, y, z] \cdot \mathbf{R}_x(-\theta_{x'})\mathbf{R}_y(-\theta_{y'})\mathbf{R}_z(-\theta_{z'}) + \vec{r}_t + \vec{d}_t \end{cases}, \quad (4)$$

where $\vec{d}_t = (d_{t,x'}, d_{t,y'}, d_{t,z'})$ represents the deviation of the center of the transducer from the origin of the OCS. After these transformations, the rotation and translation of the particle can be regarded as the rotation and translation of the transducer (or the wave function) while the particle remains stationary on the CCS, which is the basis of all the following derivations.

The linearity of the problem allows us to expand the wave function (of Eq. (4)) as a series of spherical harmonic functions using the partial wave expansion [39]. To simplify the analysis, we represent the acoustic pressure field $\hat{p}$ using the acoustic potential field $\hat{\phi}$, following relationship of $\hat{\phi} = \frac{\hat{p}}{\omega\rho_0}$i under the time-harmonic background.

$$\hat{\phi}_{\text{trans}} = \hat{\phi}_0 \sum_{n,m} a_{nm} J_n^m, \qquad (5)$$

where acoustic potential field $\hat{\phi}_{\text{trans}}$ is abbreviated from $\hat{\phi}_{\text{trans}}(\vec{r})$ at a specific position $\vec{r} = (r, \theta, \phi)$ under the CCS, and function $J_n^m \equiv j_n(kr)Y_n^m(\theta, \phi)$. $j_n(kr)$ is the spherical Bessel function of order $n$ at a position $r$ and $Y_n^m(\theta, \phi)$ is the spherical harmonic function of $n$-th order and $m$-th degree at the angular position $(\theta, \phi)$. Abbreviation $\sum_{n,m} \equiv \sum_{n=0}^{+\infty} \sum_{m=-\infty}^{+\infty}$. The expansion coefficients $a_{nm}$, or the beam-shape coefficients, can be obtained from the incident field using the orthogonality relation of the spherical harmonic functions, which is $\int_0^{2\pi}\int_0^{\pi} Y_n^m(\theta,\phi) Y_{n'}^{m'}(\theta,\phi) \sin(\theta) \, d\theta d\phi = \delta_{nn'}\delta_{mm'}$, where $\delta_{nm}$ is the Kronecker delta function. Then, the beam-shape coefficients can be evaluated by employing the orthogonality properties on Eq. (5):

$$a_{nm} = \frac{1}{\hat{\phi}_0 j_n(kR)} \int_0^{2\pi}\int_0^{\pi} \hat{\phi}_{\text{trans}}(\vec{R}) Y_n^m(\theta,\phi)^* \sin(\theta) \, d\theta d\phi, \qquad (6)$$

where $\vec{R}$ describes a spherical region in which the incident wave propagates under the CCS; the spherical region should contain the scatterer, not sound sources (i.e., $a < R < d_t$). The superscript symbol $*$ means taking conjugation of the corresponding variable. Here, the potential field $\hat{\phi}_{\text{trans}}(\vec{R}) = \frac{\hat{p}(\vec{R})}{\omega\rho_0}$i, and $\hat{p}(\vec{R})$ is given in Eq. (4).

## 2.3. Wave function of a transducer array

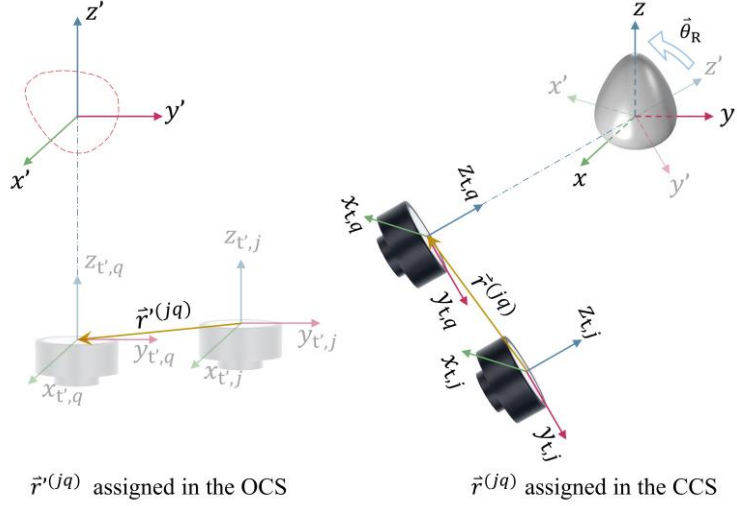

$\vec{r}'^{(jq)}$ assigned in the OCS        $\vec{r}^{(jq)}$ assigned in the CCS

**Figure 2:** Geometric description of the position relationship of transducers in the OCS and the CCS. The probe transducer (marked as $q$) and the source transducers (marked as $j$) can be linked by a relative position vector $\vec{r}^{(jq)}$ under the CCS, in which the potential field from the source transducer can be expressed in the probe transducer system $(x_{t,q}, y_{t,q}, z_{t,q})$ with the help of the translation addition theorem (Eq. (9)). Between the OCS and the CCS, a rotation transformation of Eq. (7) is needed to transform the relative position vector from $\vec{r}'^{(jq)}$ to $\vec{r}^{(jq)}$.

Figure 2 illustrates the position relationship of any two transducers in the transducer array before and after coordinate transformation from the OCS to the CCS. Here, we choose one transducer as the probe transducer with index $i = q$, and the rest are the source transducers with index $i = j$ and $j \neq q$. The position vectors $\vec{r}^{(i)}$ describe the field points located in the $(x_{t,i}, y_{t,i}, z_{t,i})$ transducer coordinates. The source transducers of the index $j$ are located by the ($q$-th) probe transducer as $\vec{r}'^{(jq)}$ described in the OCS, while it is denoted as $\vec{r}^{(jq)}$ on the CCS. The relative position vector $\vec{r}^{(jq)}$ assigned on the CCS is obtainable from the known $\vec{r}'^{(jq)}$ located in the OCS by applying a rotation transformation as

$$\vec{r}^{(jq)} = \vec{r}'^{(jq)} \cdot \mathbf{R}_x(\theta_{x'})\mathbf{R}_y(\theta_{y'})\mathbf{R}_z(\theta_{z'}). \tag{7}$$

All transducers have the same diameter $d$, while operating under different ultrasound transducer parameters $A_i e^{i\alpha_i}$ (amplitude $A_i$ and phase $\alpha_i$ of the ultrasound transducer excitation signal), $i = 1,2,\cdots,N_{\text{t}}$ with the total number of transducers $N_{\text{t}}$. Without loss of generality, we assume that wave function from all transducers follows Eq. (3) in their respective transducer coordinates, while the acoustic potential field generated by the $q$-th transducer is formulated by Eq. (5) under the CCS. Considering the transducer parameters, the potential field $\hat{\phi}_{\text{trans}}^{(q)}(\vec{r})$ (denoted as $\hat{\phi}_{\text{trans}}^{(q)}$) generated by the $q$-th transducer at position $\vec{r} = (r,\theta,\phi)$ becomes

$$\hat{\phi}_{\text{trans}}^{(q)} = \hat{\phi}_0 \sum_{n,m} a_{nm}^{(q)} J_n^m, \tag{8}$$

where the expansion coefficients $a_{nm}^{(q)} = A_q e^{i\alpha_q} \cdot a_{nm}$, namely transducer beam-shape coefficients of the $q$-th transducer.

The potential field of other source transducers is obtainable with the help of the translation addition theorem [40]. As illustrated in Fig. 2, the position vector $\vec{r}^{(j)}$ of the $j$-th transducer coordinates and the position vector $\vec{r}^{(q)}$ of the $q$-th transducer coordinates can be linked by relative position vector as $\vec{r}^{(j)} = \vec{r}^{(jq)} + \vec{r}^{(q)}$. With the help of relative position vector $\vec{r}^{(jq)}$ derived in Eq. (7), the potential field generated by the $j$-th transducer can be consistently formulated as [26]

$$\hat{\phi}_{\text{trans}}^{(j)} = \hat{\phi}_0 \sum_{n,m} \tilde{a}_{nm}^{(jq)} J_n^m, \tag{9}$$

where expansion coefficients $\tilde{a}_{nm}^{(jq)} = \sum_{v,\mu} a_{v\mu}^{(j)} S_{v,n}^{\mu,m\,(1)}(k\vec{r}^{(jq)})$, defined as the transformation beam-shape coefficients of the $j$-th transducer. $S_{v,n}^{\mu,m\,(1)}(k\vec{r}^{(jq)})$ is the separation transform matrix of the first kind [40], used to transform the information from the $j$-th transducer coordinate system to the $q$-th transducer coordinate system. The linearity of the problem allows us to represent the potential field of the whole transducer array as a summation of the contributions from all transducers: $\hat{\phi}_{\text{ex}} = \hat{\phi}_0 \sum_{i=1}^{N_{\text{t}}} \hat{\phi}_{\text{trans}}^{(i)}$. Considering the potential fields generated by the probe transducer in Eq. (8) and the source transducers in Eq. (9), we arrive

$$\hat{\phi}_{\text{ex}} = \hat{\phi}_0 \sum_{n,m} \tilde{a}_{nm} J_n^m, \tag{10}$$

where the external potential field $\hat{\phi}_{\text{ex}}$ is abbreviated from $\hat{\phi}_{\text{ex}}(\vec{r})$ at a specific position $\vec{r} = (r,\theta,\phi)$ under the CCS, and the expansion coefficients $\tilde{a}_{nm} = a_{nm}^{(q)} +$ $\sum_{j \neq q} \tilde{a}_{nm}^{(jq)}$, defined as the beam-shape coefficients of the transducer array. Abbreviation $\sum_{j \neq q} \equiv \sum_{i=1, i \neq q}^{N_t}$.

## 2.4. The Helmholtz wave equation

After establishing the overall framework of the external wavefield transformation from the OCS to the CCS, we can further estimate the scattered potential field reflected by an irregular scatterer. In the source-free regions of the physical space, the total potential field satisfies the Helmholtz wave equation

$$(\nabla^2 + k^2)\hat{\phi} = 0, \tag{11}$$

where $\nabla^2$ is the Laplacian operator. The total potential field is contributed by the external potential field $\hat{\phi}_{ex}$ and the scattering potential field reflected by the scatterer $\hat{\phi}_{sc}(\vec{r})$ (denoted as $\hat{\phi}_{sc}$)

$$\hat{\phi} = \hat{\phi}_{ex} + \hat{\phi}_{sc}, \tag{12}$$

Here, the linearity of the problem allows us to represent the scattering potential field as a series of spherical harmonics function [39]

$$\hat{\phi}_{sc} = \hat{\phi}_0 \sum_{n,m} s_{nm} \tilde{a}_{nm} H_n^m. \tag{13}$$

The scalar scattering coefficients, $s_{nm}$, almost depend on the boundary conditions. Function $H_n^m \equiv h_n(kr) Y_n^m(\theta, \phi)$. $h_n(kr)$ is the Hankel function of the first kind at position $r$. Dirichlet or Neumann boundary conditions require that the total acoustic pressure or the normal particle velocity vanishes on the surface of the scatterer. For the particles under consideration, this can be stated, respectively, as:

$$[\hat{\phi}_{ex}(\vec{\Omega}) + \hat{\phi}_{sc}(\vec{\Omega})] = 0, \tag{14a}$$

$$\vec{n} \cdot \nabla [\hat{\phi}_{ex}(\vec{\Omega}) + \hat{\phi}_{sc}(\vec{\Omega})] = 0. \tag{14b}$$

where $\vec{n}$ is the outer normal vector to the scatterer surface $\vec{\Omega}$.

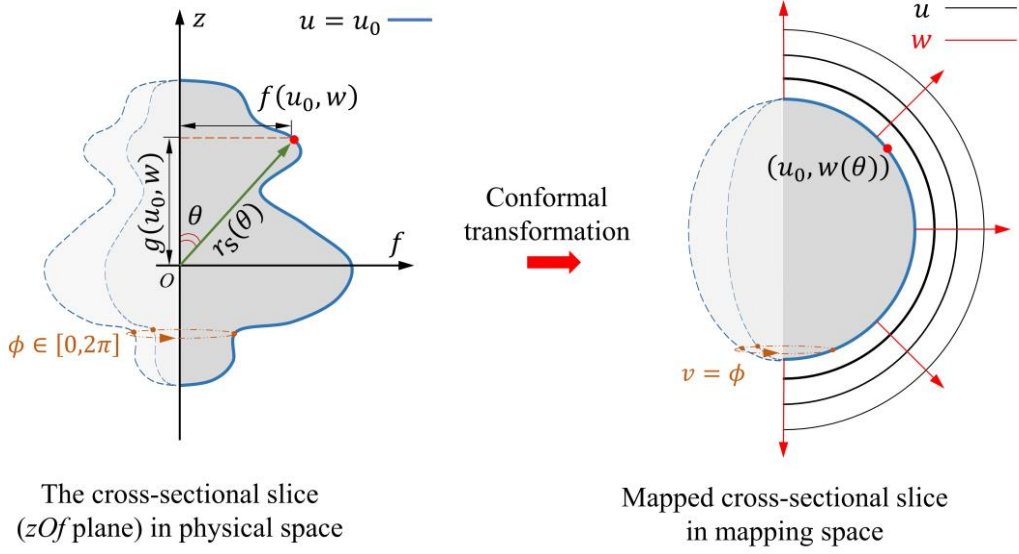

**Figure 3:** Conformal transformation mapping of an axisymmetric particle in physical space to a sphere in mapping space. The particle is symmetric about the $z$-axis. The particle surface in the mapping coordinate $(u, w, v)$ system can be described by the new radial coordinate $u = u_0$, independent with the new polar angular coordinate, $w$, and the azimuthal angular coordinate, $v$. On the $zOf$ slice plane, defined as arbitrary cross-sectional plane along the symmetric axis ($z$-axis), the radial coordinate satisfies $r_s(u_0, w) = \sqrt{f^2(u_0, w) + g^2(u_0, w)}$ in the real-system, while $r_s(\theta)e^{i\theta} = g(u_0, w) + f(u_0, w) \cdot i$ in the complex-system (the azimuthal angular coordinates, $\phi$ or $v$, are not involved for the axisymmetric reason). The mapping functions $g(u_0, w)$ and $f(u_0, w)$ are introduced to connect the physical space and the mapping space.

To analytically evaluate the scattered fields (i.e., the scalar scattering coefficients $s_{nm}$) for any given incident wave, we must solve the Helmholtz wave equation (Eq. (11)) subject to the irregular boundary conditions along the particle surface (Eq. (14)). However, due to the boundary surfaces $\vec{\Omega}$ that are generally inseparable and thus incompatible with the method of separation of variables, it is impractical to establish an analytical solution to the Helmholtz wave equation. We attempt to map the $(r, \theta, \phi)$ physical space inhabited by the irregular scatterer to a new quasi-spherical coordinate that is denoted as $(u, w, v)$ system, in where the locus of all points of the scatterer boundary for the new radial coordinate, $u$, is equal to a constant ($u = u_0 = 0$). The new polar angular coordinate of the mapping coordinate system, $w$, corresponds to the spherical polar angular coordinate, $\theta$. Since the body is symmetric along the $z$-axis, the new azimuthal angular coordinate, $v$, remains identically with the spherical azimuthal angular coordinate $\phi$, varied from 0 to $2\pi$. Figure 3 shows the geometry and mapping information of an axisymmetric particle on different coordinate systems. The center of mass of the irregular body is set to coincide with the origin of the physical coordinate systems. The $zOf$ plane defines as a two-dimensional physical space where the azimuthal angular variable $\phi$ is a constant. Although there are infinite $zOf$ planes for different azimuthal angular variables, the cross-sectional slice of an axisymmetric object on any $zOf$ plane is identical. Let us consider a complex mapping function $r_s(\theta)e^{i\theta} = M(u + w \cdot i)$ [30][31], which maps an irregular cross-sectional slice $r_s(\theta)$ described on the $zOf$ physical space to a circle on $(u, w)$ space, according to

$$M(u + w \cdot i) = g(u, w) + f(u, w) \cdot i, \tag{15}$$

$$\begin{cases} g(u, w) = c_{-1}e^u \cos(w) + \sum_{n=0}^{\infty} c_n e^{-nu} \cos(w) \\ f(u, w) = c_{-1}e^u \sin(w) - \sum_{n=0}^{\infty} c_n e^{-nu} \sin(w) \end{cases},$$

where $c_n, n = -1, 0, 1, \cdots, \infty$ are the mapping coefficients. Under the complex plane, the coordinates in physical and mapping spaces should satisfy

$$\begin{cases} r_s(u, w) = \sqrt{f^2(u, w) + g^2(u, w)} \\ \theta(u, w) = \cos^{-1}\left(g(u, w)/r_s(u, w)\right) \end{cases}. \tag{16}$$

Then, a set of mapping coefficients $c_n$ can be determined by equating the slice function (Eq. (16)) to the mapping functions (Eq. (15)) on the scatterer surface (i.e., $u = u_0$). Detailed processes to estimate the mapping coefficients can be found in Appendix A. Note that the shape of the boundary of any cross-sectional slice for any specified azimuthal angle $\phi = v \in [0, 2\pi]$ is identical. The scatterer can be regarded as a cross-sectional slice rotating along the azimuthal angular coordinate for a $2\pi$ period, and a three-dimensional conformal mapping is achievable.

A well-known result is that under a conformal transformation mapping, the Helmholtz wave equation (Eq. (11)) takes a new form in the $(u, w)$-plane [41][42] given by

$$\left(\nabla^2 + k^2 \Im(u,w)\right)\hat{\phi}(u,w,v) = 0, \tag{17}$$

where $\Im(u,w)$ is the Jacobian of the transformation from $(r_s(\theta),\theta)$ system to $(u,w)$ system. Evidently, if $\hat{\phi}$ is any solution of the Helmholtz wave equation (in Eq. (11)), in the spherical coordinate system, then $\hat{\phi}(u,w,v)$ is a solution of conformal mapping coordinates, Eq. (17) [43]. Following the established results, we can formulate the external and scattered potential fields on the mapping coordinates by transforming Eqs. (10) and (13) as

$$\hat{\phi}_{\text{ex}}(u,w,v) = \hat{\phi}_0 \sum_{n,m} \tilde{a}_{nm} J_n^m(u,w,v), \tag{18}$$

and

$$\hat{\phi}_{\text{sc}}(u,w,v) = \hat{\phi}_0 \sum_{n,m} s_{nm} \tilde{a}_{nm} H_n^m(u,w,v), \tag{19}$$

where abbreviations $J_n^m(u,w,v) \equiv j_n(kr(u,w))Y_n^m(\theta(u,w),v)$ and $H_n^m(u,w,v) \equiv h_n(kr(u,w))Y_n^m(\theta(u,w),v)$. The quantities $r(u,w)$ and $\theta(u,w)$ can be determined by Eq. (16). A summation of Eqs. (18) and (19) gives the total potential field in terms of the new coordinates $(u,w,v)$, which also is the solution of Eq. (17).

An equivalent representation of Dirichlet and Neumann conditions (Eq. (14)) are then becomes

$$[\hat{\phi}_{\text{ex}}(u_0,w,v) + \hat{\phi}_{\text{sc}}(u_0,w,v)] = 0, \tag{20a}$$

$$\vec{n} \cdot \nabla[\hat{\phi}_{\text{ex}}(u_0,w,v) + \hat{\phi}_{\text{sc}}(u_0,w,v)] = 0. \tag{20b}$$

To effectively leverage these conditions, we insert Eqs. (18) and (19) into Eq. (20), and multiply the results by a set of spherical angular eigenfunctions, $\psi_{n'}^{m'}(w,v) = P_{n'}^{m'}(\cos(w))\sin(w)e^{-m'v\mathrm{i}}$ (or Eq. (B.6)); the derivations are further independent of coordinates by integrating the over the range of $w$ and $v$, yielding

$$\sum_{n=0}^{\infty} \tilde{a}_{nm'} \Gamma_n^{n',m'} + \sum_{n=0}^{\infty} s_{nm'} \tilde{a}_{nm'} \Lambda_n^{n',m'} = 0, \tag{21}$$
$$(n' = 0,1,\cdots,\infty; m' = -\infty,\cdots,0,\cdots,\infty)$$

and

$$\sum_{n=0}^{\infty} \tilde{a}_{nm'} \Gamma_{n,u}^{n',m'} + \sum_{n=0}^{\infty} s_{nm'} \tilde{a}_{nm'} \Lambda_{n,u}^{n',m'} = 0, \tag{22}$$
$$(n' = 0,1,\cdots,\infty; m' = -\infty,\cdots,0,\cdots,\infty)$$

where the structural functions, $\Gamma_n^{n',m'}$ and $\Lambda_n^{n',m'}$, and their partial derivatives of new radial coordinate, $\Gamma_{n,u}^{n',m'}$ and $\Lambda_{n,u}^{n',m'}$, are listed in Eqs. (B.9) and (B.13). Note that a complete derivation of the above processes is written in Appendix B.

In order to solve the problem, the infinite summations in Eqs. (18), (19), (21), and (22) have been truncated to $N$ terms to facilitate their computation, which limits the summations from $\sum_{n,m} \equiv \sum_{n=0}^{+\infty} \sum_{m=-\infty}^{+\infty}$ to $\sum_{n,m} \equiv \sum_{n=0}^{N} \sum_{m=-N}^{N}$. It can be seen that the total number of unknown variables of the scalar scattering coefficients $s_{nm}$ in Eq. (19) includes $(N+1) \times (2N+1)$ elements. Matrices (21) and (22) offer a set of $N+1$ equations for each fixed $m' \in [-N, N]$, and totally a set of $(N+1) \times (2N+1)$ equations for Dirichlet (sound-soft) and Neumann (sound-hard) boundary conditions, respectively. Hence, the scalar scattering coefficients and thus the scattered potential field is determined by solving these linear equations. A method to solve the equation system is given in Appendix C.

## 2.5. Acoustic radiation force and torque

The acoustic radiation force and torque acting on a spherical object excited by a transducer array under the mapping coordinate system are given by [24][25][26]

$$F_{\text{rad},x} = -\frac{\hat{\phi}_0^2 \rho_0}{4} \text{Re}\left[ \text{i} \cdot \sum_{n,m} \tilde{a}_{nm}(1 + s_{nm})\left(\mathcal{A}_{n+1}^{m+1} b_{n+1,m+1}^* - \mathcal{B}_{n-1}^{m+1} b_{n-1,m+1}^* + \mathcal{C}_{n+1}^{m-1} b_{n+1,m-1}^* - \mathcal{D}_{n-1}^{m-1} b_{n-1,m-1}^* \right) \right], \tag{23}$$

$$F_{\text{rad},y} = -\frac{\hat{\phi}_0^2 \rho_0}{4} \text{Re}\left[\sum_{n,m} \tilde{a}_{nm}(1 + s_{nm})\left(\mathcal{A}_{n+1}^{m+1} b_{n+1,m+1}^* - \mathcal{B}_{n-1}^{m+1} b_{n-1,m+1}^* - \mathcal{C}_{n+1}^{m-1} b_{n+1,m-1}^* + \mathcal{D}_{n-1}^{m-1} b_{n-1,m-1}^*\right)\right],$$

$$F_{\text{rad},z} = -\frac{\hat{\phi}_0^2 \rho_0}{2} \text{Re}\left[i \cdot \sum_{n,m} \tilde{a}_{nm}(1 + s_{nm})\left(\mathcal{E}_{n+1}^{m} b_{n+1,m}^* - \mathcal{F}_{n-1}^{m} b_{n-1,m}^*\right)\right],$$

and

$$T_{\text{rad},x} = -\frac{\hat{\phi}_0^2 \rho_0}{4k} \text{Re}\left[\sum_{n,m} \tilde{a}_{nm}(1 + s_{nm})\left(\mathcal{G}_n^{m} b_{n,m+1}^* + \mathcal{G}_n^{-m} b_{n,m-1}^*\right)\right], \quad (24)$$

$$T_{\text{rad},y} = -\frac{\hat{\phi}_0^2 \rho_0}{4k} \text{Re}\left[i \cdot \sum_{n,m} \tilde{a}_{nm}(1 + s_{nm})\left(\mathcal{G}_n^{m} b_{n,m+1}^* - \mathcal{G}_n^{-m} b_{n,m-1}^*\right)\right],$$

$$T_{\text{rad},z} = -\frac{\hat{\phi}_0^2 \rho_0}{2k} \text{Re}\left[\sum_{n,m} \tilde{a}_{nm}(1 + s_{nm}) m b_{n,m+1}^*\right],$$

where abbreviation $b_{nm} = \tilde{a}_{nm} \cdot s_{nm}$ and symbol Re means taking the real part of the expression. The weighting coefficients $\mathcal{A}_n^m = -\mathcal{C}_n^m = -\sqrt{\frac{(n+m-1)(n+m)}{(2n-1)(2n+1)}}$, $\mathcal{B}_n^m = -\mathcal{D}_n^m = \sqrt{\frac{(n-m+2)(n-m+1)}{(2n+1)(2n+3)}}$, $\mathcal{E}_n^m = \mathcal{F}_n^m = \sqrt{\frac{(n-m)(n+m)}{(2n-1)(2n+1)}}$, and $\mathcal{G}_n^m = \sqrt{(n-m)(n+m+1)}$.

Consider that the new mapping coordinate system becomes a spherical coordinate system when the new radial coordinate tends to be infinite, $u \to +\infty$, and thus the scalar scattering coefficients, $s_{nm}$, solved in matrices (21) and (22), are acceptable to describe the scattered field reflected by the irregular particle under the physical space in the limit of great distances from the scatterer. Consequently, the acoustic radiation force and torque that are evaluated using the far-field data can be asymptotically formulated using Eqs. (23) and (24) without performing an inverse mapping from the mapping space to the physical space.

Note that the radiation force and torque are estimated on the CCS. Another rotation transformation is required to transform the radiation force and torque from the CCS to the OCS using Eq. (2) (i.e., step two illustrated in Fig. 1):

$$[F_{\text{rad},x'}, F_{\text{rad},y'}, F_{\text{rad},z'}] \quad (25)$$
$$= [F_{\text{rad},x}, F_{\text{rad},y}, F_{\text{rad},z}] \cdot \mathbf{R}_z^{-1}(\theta_{z'})\mathbf{R}_y^{-1}(\theta_{y'})\mathbf{R}_x^{-1}(\theta_{x'}),$$

and

$$[T_{\text{rad},x'}, T_{\text{rad},y'}, T_{\text{rad},z'}] \quad (26)$$
$$= [T_{\text{rad},x}, T_{\text{rad},y}, T_{\text{rad},z}] \cdot \mathbf{R}_z^{-1}(\theta_{z'})\mathbf{R}_y^{-1}(\theta_{y'})\mathbf{R}_x^{-1}(\theta_{x'}).$$

In this way, the radiation force and torque acting upon an irregular particle with arbitrary orientation from a transducer array can be obtained, which is the basis for later discussion of the translational and rotational dynamics, i.e., prediction of the acoustophoresis of an irregular particle.

## 2.6. Dynamic manipulation

When a particle is placed above an ultrasound transducer array, it mainly experiences radiation force and torque that cause translational and rotational motions, the drag force $\vec{F}_{\text{drag}}$ and drag torque $\vec{T}_{\text{drag}}$ due to the viscous stresses and shear stresses on the particle surface, and its gravity $\vec{F}_G$. The translational and rotational movements of the particle are then described *via* the equations of motion as

$$m_p \frac{d\vec{u}_p}{dt} = \vec{F}_{\text{rad}} + \vec{F}_{\text{drag}} + \vec{F}_G, \quad (27)$$

and

$$I_p \frac{d\vec{\omega}_p}{dt} = \vec{T}_{\text{rad}} + \vec{T}_{\text{drag}}. \quad (28)$$

where $m_p$ is the mass of the particle and $I_p$ is the moment of inertia of the particle. $\vec{u}_p$ and $\vec{\omega}_p$ are translational particle velocity and angular velocity about its center of mass, respectively. The drag force and torque are approximately evaluated using classical formulas [35][36] as

$$\vec{F}_{\text{drag}} = 6\pi a \eta \vec{u}, \quad (29)$$

and

$$\vec{T}_{\text{drag}} = 8\pi a^3 \eta \vec{\omega}, \quad (30)$$

where $a$ is averaged radius of the particle and $\eta$ is the dynamic viscosity of the host fluid. The velocity $\vec{u}$ and angular velocity $\vec{\omega}$ are based on the relative velocity of the particle with respect to the background fluid. In our case, the fluid is assumed to be at rest, thus $\vec{u} = -\vec{u}_p$ and $\vec{\omega} = -\vec{\omega}_p$. Consider that in many acoustofluidic scenarios, the inertia of the particle can be neglected since the characteristic time of acceleration is small in comparison to the time scale of the motion ($\Delta t$) of the particles [44]. Based on this assumption, we can identify the translational and rotational trajectories, i.e., particle acoustophoresis, with the particle velocity $\vec{u}_p$ and particle angular velocity $\vec{\omega}_p$ by

$$\vec{u}_p = \frac{\vec{F}_{rad} + \vec{F}_G}{6\pi a \eta}, \tag{31}$$

and

$$\vec{\omega}_p = \frac{\vec{T}_{rad}}{8\pi a^3 \eta}. \tag{32}$$

In this way, we have computed the solution of the dynamics problem of a non-spherical particle under its weight, encompassing viscous drag, and acoustic radiation forces and torques. The time-dependent system was solved based on a simple time accumulation method by providing an initial position for the particle. The dynamics were simulated for 2 s with a time step of $\Delta t = 0.1$ ms. The acoustic radiation forces and torques are recalculated for the new position and orientation. Specifically, at each time step ($\Delta t$), we need to determine the states of the particle, i.e., obtain the particle displacement and rotation angle using particle velocity and angular velocity in current time step through $\vec{r}_t = \vec{r}_t + \Delta \vec{r}_t$ with $\Delta \vec{r}_t = \vec{u}_p \Delta t$, and $\vec{\theta}_{rot} = \vec{\theta}_{rot} + \Delta \vec{\theta}_{rot}$ with $\Delta \vec{\theta}_{rot} = \vec{\omega}_p \Delta t$. The new position and orientation are then used to link the OCS and the CCS for the next time step calculation.

# 3. Results and discussion

## 3.1. Model preparation

We need to impose a truncation number of partial wave series, $N$, in the number of modes entering the computations of acoustic radiation force in Eq. (23) and acoustic radiation torque in Eq. (24). Although it is able to further improve the prediction accuracy by enforcing as many modes as possible to enter the computations, the truncation number $N = ka + 6 \approx 8$ [45] is basically enough to converge the semi-analytical radiation force and torque to the corresponding full three-dimensional numerical solutions. Note that for $N = 8$, it takes about 5 s in evaluating a set of radiation force and torque simultaneously (in PC with CPU: Intel i7-6700HQ 2.6 GHz, and Maximum memory usage: 16 GB).

In order to evaluate the beam-shape coefficients $a_{nm}$ in Eq. (6), we need to specify a spherical space with a radius of R, in which the potential field can be approximately described by the model expansion series as given in Eq. (5). A larger R means that the approximated space has been wider. If the truncation number $N = 8$ is fixed (i.e., the number of the beam-shape coefficients in Eq. (6) is fixed), a larger R indicates that the same number of beam-shape coefficients are used to approximate a wider space, which inevitably impairs the predicted precision, even distorts the prediction results. In contrast, for a smaller R, it is equivalent to using the same number of beam-shape coefficients to approximate a smaller space, which may lead to overfitting of the potential field. Here, we introduce the radial intensity to quantify the approximation:

$$I_\mathrm{r} = \frac{1}{2}\mathrm{Re}(\hat{p}\hat{v}_\mathrm{r}^*), \tag{33}$$

where the pressure amplitude $\hat{p} = -\omega\rho_0\hat{\phi}\mathrm{i}$, and the radial velocity amplitude $\hat{v}_\mathrm{r} = -\frac{\partial\hat{\phi}}{\partial r}$.

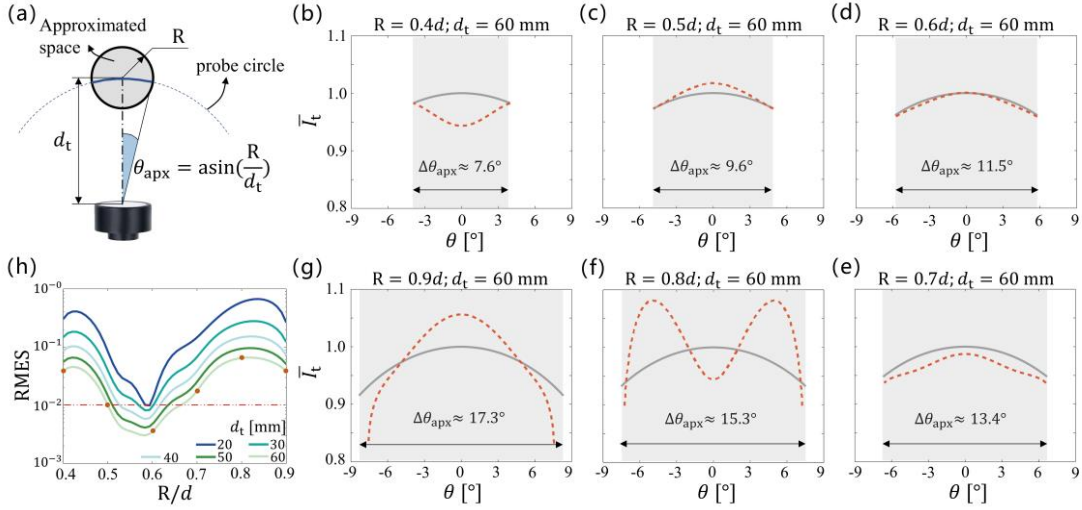

**Figure 4:** Root-mean-square error (RMSE) of the normalized radial intensity ($\bar{I}_r$) between the modal expansion approximations and the theoretical results along the probe arc ( $\theta \in [-\theta_{apx}, \theta_{apx}]$ ). (a) The geometric relationship among the approximated space, the probe circle, and the probe arc. The approximated space is a spherical domain with a radius of R, and its center is consistent with the mass center of the contained scatterer. The radius of the probe circle is $d_t$ and its center is located at the center of the transducer surface. The intersection of the approximated space and the probe circle is defined as the probe arc. (b)-(g) Visualization of the differences of the normalized radial intensity along the probe arc. The solid and dashed red curves denote the results based on the modal expansion series and theoretical solutions, respectively. (h) RMSE of $\bar{I}_r$ along the probe arc as a function of R and $d_t$. The brown dots represent the cases given in (b) to (g).

Figure 4 shows the directivity of the normalized radial intensity $\bar{I}_r = \frac{I_r}{\max(I_r)}$ along a probe arc. The probe arc is a segment of the probe circle in the approximated space, as illustrated in Fig. 4(a). The radius of the probe circle is set to $d_t$, consistent with the distance between the scatterer and the circular radiator. The radius of the spherical approximated space is R, located right above the sound source. Here, the radius R is ranged from $0.4d$ to $0.9d$ with the transducer diameter of $d = 10$ mm, while $d_t$ is taken as 20 mm to 60 mm with an interval of 10 mm. The directivity of normalized radial intensity based on the modal expansion series is compared with the theoretical counterparts (derived from Eq. (3)), as plotted in Figs. 4(b) to 4(g). The difference is almost invisible in Fig. 4(d). Note that we only present the polar angle ranging from $-\theta_{\text{apx}}$ to $\theta_{\text{apx}}$ ($\theta_{\text{apx}} = \arcsin\left(\frac{R}{d_t}\right)$) since the beam-shape coefficients and the potential field given in Eqs. (5) and (6) are valid inside the approximated space. In order to quantify the differences, we calculate the root-mean-square error (RMSE), which is illustrated in Fig. 4(h). It can be found that the trends are basically the same for different $d_t$, while the errors vary significantly for different R. The RMSE becomes minimum ($\leq 1\%$) when $R \approx 0.6d$; larger or smaller R increases the errors. Hence, the radius of the approximated space is set to $R = 0.6d$ in later computations.

**Table 1:** Mapping coefficients for different axisymmetric particles in calculations. Parameter $a$ is the averaged radius of the axisymmetric geometries.

| Mapping coefficients | Sphere | Ellipsoid | Cone | Diamond |
|---|---|---|---|---|
| $c_{-1}$ | $a$ | $a$ | $a$ | $a$ |
| $c_1$ | 0 | $a/5$ | 0 | 0 |
| $c_2$ | 0 | 0 | $a/8$ | 0 |
| $c_3$ | 0 | 0 | 0 | $a/10$ |
| $c_n, n \geq 0$ and $n \neq 1,2,3$ | 0 | 0 | 0 | 0 |

Furthermore, we need to prepare the mapping coefficients $c_n$. Although a general solution for arbitrary geometries based on series expansions is available in Appendix A, many practical geometries do not require such a comprehensive procedure. For the typical geometries, including ellipsoid, triangular cone, diamond, and sphere, we give the mapping coefficients, $c_n$, in Tab. 1. The geometric differences are captured by different combinations of the mapping coefficients $c_n$, while the geometric size can be stretched by adjusting the averaged radius $a$.

### 3.2. Validation and discussion

In the following subsections, full three-dimensional finite element simulations are conducted with COMSOL Multiphysics 5.5 to provide validations for the proposed analytical techniques as prescribed in Eqs. (25) and (26).

A rectangular region ($24a \times 24a \times 18a$) is defined as the simulation domain. The center of mass of the irregular particles and the center of the simulation domain are both placed at the origin of the Cartesian coordinate system. A spherical surface with a radius of $\sim 0.7d$ is defined to divide the whole simulation domain into two sub-domains, a finer mesh domain and a coarser mesh domain inside and outside the surface, respectively. We set another numerical integration surface with a radius of $R \approx 0.5d$ inside the finer mesh domain as the integration surface to numerically evaluate the radiation force and torque by inserting the sound pressure and particle velocity into Eqs. (D.4) and (D.5). For solid particles levitated in the air, we usually apply Neumann (or sound-hard) boundary conditions to the particle surface in numerical simulations, which correspond to the scalar scattering coefficients solved by Eq. (B.12) used in our method. To make the wavefield radiated from a circular surface in the simulation approximately consistent with that given in Eq. (3), we can set the circular radiator with a radial vibrated velocity of $\hat{v}_0$, which is the same as that used in Eq. (3). In this way, the circular surface can radiate a wavefield approximately expressed by Eq. (3) in the far-field region. The Sommerfeld radiation condition is required to eliminate the reflected wave, achievable by applying the perfect matched layer (PML) surrounding the simulation domain. Following the above considerations, we summarize the simulational parameters in Tab. 2. The detailed information can refer to the numerical model in the Supplemental COMSOL File.

**Table 2:** General parameters used in the finite-element simulations in COMSOL at room temperature and pressure. Note that the geometry of different particles is formulated in Eq. (16), where the mapping functions and mapping coefficients are given in Eq. (15) and Tab. 1, respectively.

| Parameter | Value |
| --- | --- |
| Average radius of bodies ($a$) | 2 mm |
| Transducer diameter ($d$) | 10 mm |
| Interdistance ($d_t$) | 20 mm |
| Density (air $\rho_0$) | 1.224 kg/m$^3$ |
| Speed of sound (air $c_s$) | 340 m/s |
| Radial velocity ($\hat{v}_0$) | 1.5 m/s |
| Incidence polar angle ($\theta_{\text{inc}}$) | 0°, 30°, 60°, 90° |
| Frequency of external wave ($f_0$) | 40000 Hz |
| Wavelength ($\lambda$) | $c_s / f_0$ |
| Cubic simulational domain | $24a \times 24a \times 18a$ |
| Radius of integrating surface R | $0.5d$ |
| Radius of finer mesh domain | $\sim 0.7d$ |
| Maximum element size (finer mesh domain) | $\lambda / 60$ |
| Maximum element size (coarser mesh domain) | $\lambda / 6$ |
| PML depth | $\lambda / 2$ |
| CPU | Intel i7-6700HQ 2.6 GHz |
| Operating system | Windows 10 |
| Maximum memory usage | ~ 16 GB |
| Computational time per case | 10 ~ 20 mins |

The theoretical evaluations of the acoustic radiation force and torque using Eqs. (25) and (26) compared with the numerical calculations using Eqs. (D.4) and (D.5) based on FEM results are given in Fig. 5. The radiation force and torque are completely validated when the scatterer rotates along $x'$-axis in different phase distributions. For axisymmetric reasons, the radiation force along $x'$-axis, $F_{\text{rad},x'}$, and the acoustic radiation torque along $y'$- and $z'$-axes, $T_{\text{rad},y'}$ and $T_{\text{rad},z'}$, are significantly weaker than the values on other directions

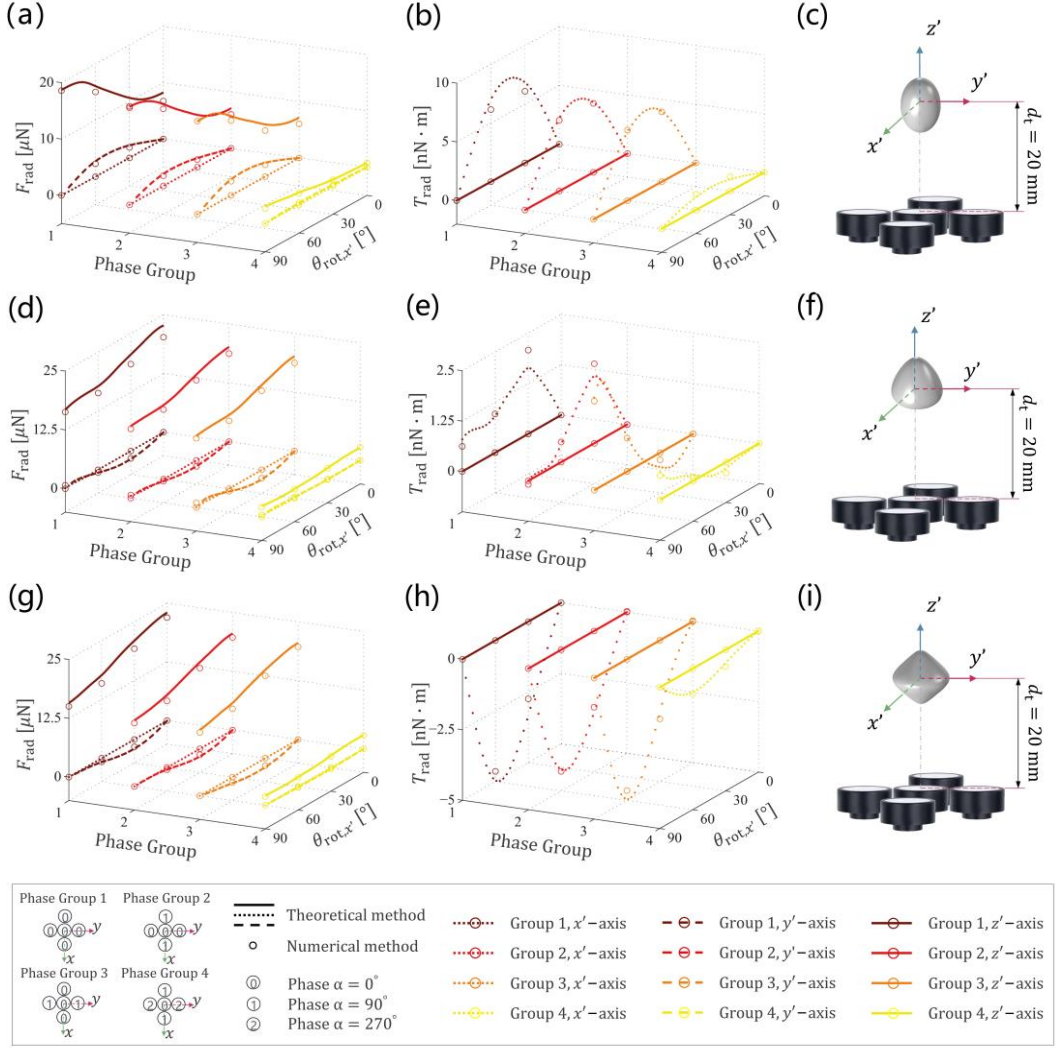

**Figure 5:** Comparisons of the radiation force $\vec{F}_{\text{rad}}$ and torque $\vec{T}_{\text{rad}}$ acting on the particles with different geometric features (average radius of $a = 2$ mm) based on the semi-analytical expansion series (Eqs. (25) and (26)) and the FEM results in a five transducer system. The radiation force and torque are plotted as a function of the rotation angle along the $x'$-axis, $\theta_{\text{rot},x'}$, under different phase distributions. Sub-figures (a), (d), and (g) display the predictions of the radiation force, while (b), (e), and (h) are of the radiation torque for (c) an ellipsoid, (f) a cone, and (i) a diamond, respectively. The circle marks represent the results based on the full three-dimensional FEM. In contrast, the dot, dashed, and solid curves mean the data collected based on the semi-analytical method along the $x'$-, the $y'$-, and the $z'$-axes, respectively. The radial velocity of the transducers are all set to $\hat{v}_0 = 1.5$ m/s, while four groups of phase distributions are applied to the transducer array.

We observe that the acoustic radiation force and torque between our method and the FEM results are almost perfectly matched. What needs to be emphasized is that we deliberately limit the height of the simulation domain in $18a = 36$ mm to reduce the number of mesh elements, which greatly saves simulation time. However, this means that the interdistance between the scatterers and the transducer array is relatively small ($d_\mathrm{t} = 20$ mm), and thus the wavefield around the scatterers does not meet the far-field requirements, which compromises the accuracy of using Eq. (3) to describe the wavefield. As a result, there are still some perceivable discrepancies in Fig. 5. It is also worth emphasizing that the computational time of each numerical simulation will take 10 to 20 minutes, which is much higher than the computational cost in our method (on the order of seconds). This computational efficiency allows us to predict the translational and rotational dynamics of the non-spherical particles levitated above a transducer array.

## 3.3. Acoustophoresis

In this section, the time-dependent system described by Eqs. (27) and (28) is solved. The non-spherical scatterers are all considered as expanded polystyrene (EPS) particles with a density of $\rho_\mathrm{p} = 15$ kg/m$^3$. In this case, the gravity can be calculated by $F_\mathrm{G} \approx \frac{4}{3}\pi a^3 \rho_\mathrm{p} g$, where $g$ is the acceleration of gravity. These particles are placed in position $\vec{r}_\mathrm{t} = (2,2,0)$ mm of the OCS, while the center of the transducer array is located at right below the origin, i.e., $\vec{d}_\mathrm{t} = (0,0,60)$ mm. The symmetric axis of these particles is initially set to coincide with the $z'$-axis, that is $\vec{\theta}_\mathrm{R} = (0,0,0)$. As an example, we assume that all transducers are operated in phase ($\alpha_i = 0; i = 1, \cdots, 9$). No further mention, other parameters used in the computations remain the same as those listed in Tab. 2. We start the predictions from $t = 0$ s with a time interval of $\Delta t = 0.1$ ms and end the predictions when the changes of the positions and the rotation angles among two adjacent time steps are less than 5 %; the dynamic trajectories of different particles are shown in Figs. 6 to 9.

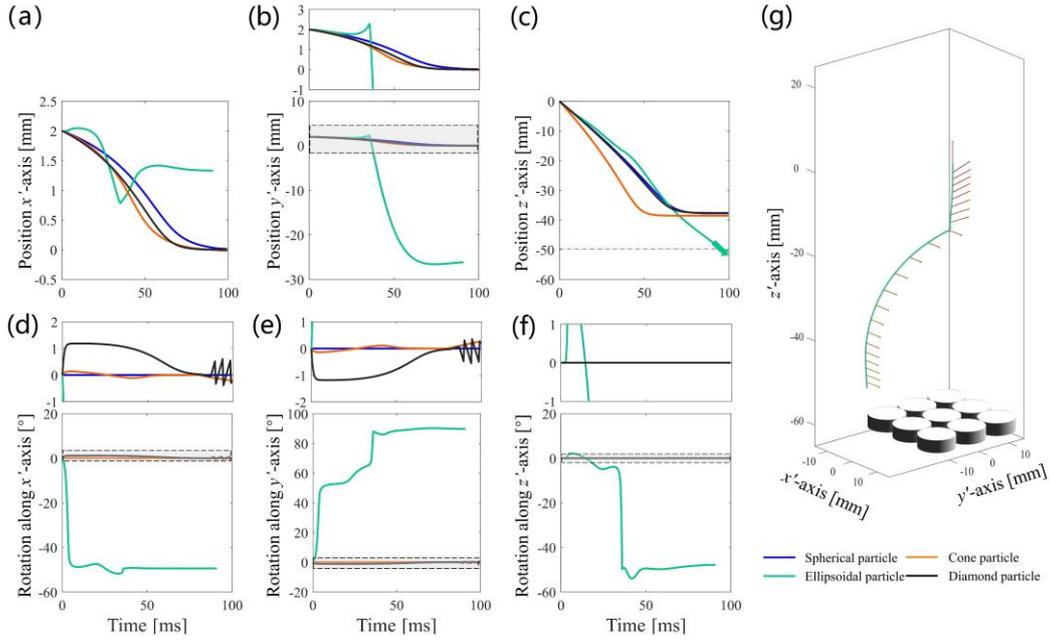

**Figure 6:** Translational and rotational dynamics of different particles with the same averaged radius of 0.5 mm. The translational trajectories of different particles at different moments along (a) $x'$-, (b) $y'$-, (c) $z'$-axes. The rotational angles of different particles at different moments along (d) $x'$-, (e) $y'$-, (f) $z'$-axes. If necessary, the gray regions are zoomed in for more details of the corresponding figures. (g) Three-dimensional translational trajectory (solid line) and particle orientation (arrow). The arrows represent the symmetric axis of the particle, while the color of the arrow is used to represent the increase of time (red-yellow color spectrum). The time intervals represented by any adjacent arrows are the same. These are 20 arrows showing the position and orientation of the particles from the start of the calculation (red arrow) to the end of the calculation (yellow arrow). Note that only the trajectory and orientation of the ellipsoidal particle are visualized, as the trajectory and orientation for other particles are not significantly different.

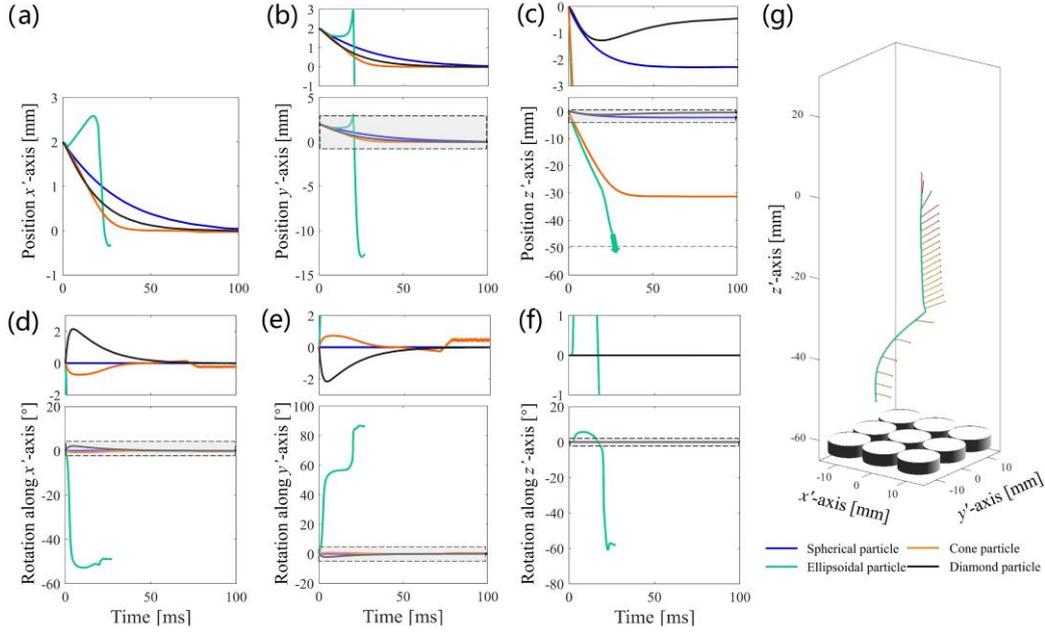

**Figure 7:** The same as in Fig. 6, but increase the averaged radius to 1 mm. Note that only the trajectory and orientation of the ellipsoidal particle are visualized in (g), as the trajectory and orientation for other particles are not significantly different.

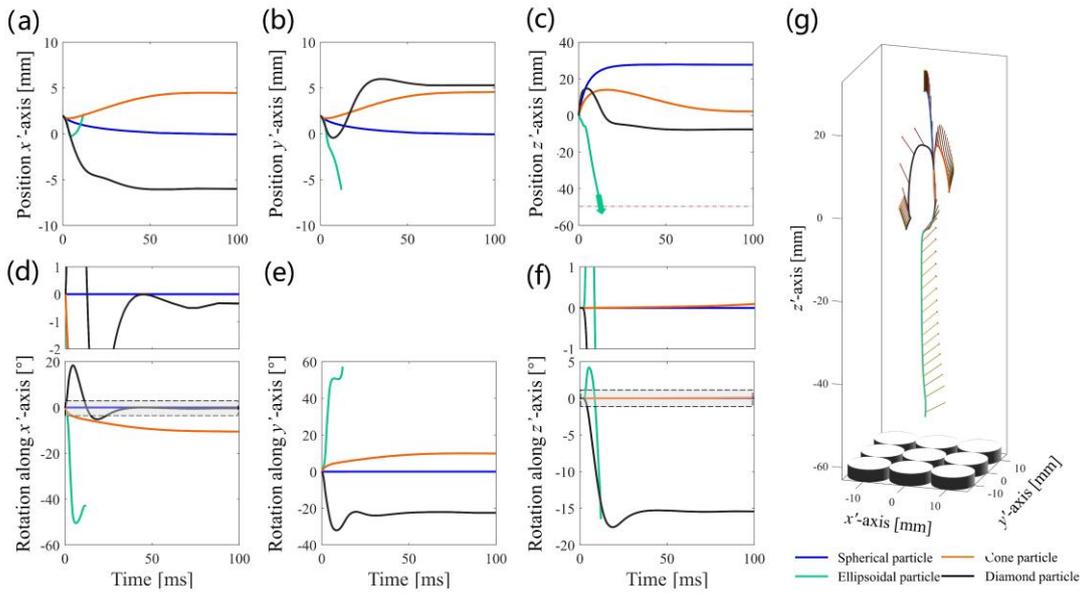

**Figure 8:** The same as in Fig. 6, but increase the averaged radius to 2 mm.

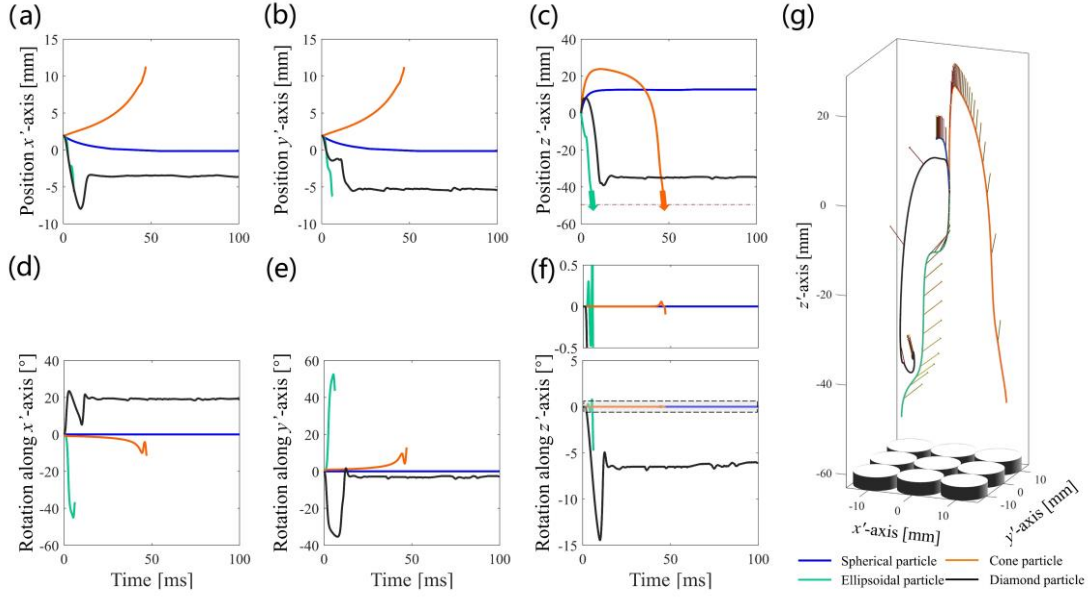

**Figure 9:** The same as in Fig. 6, but increase the averaged radius to 3 mm.

Consider that Eq. (3) is used to describe the wavefield radiated from a circular oscillator in the far-field [38]. When the particles move down 0.05 m (i.e., $d_t = 0.01$ m), we stop the calculations. This critical state is highlighted by the dashed red lines in sub-figure (c) of Figs. 6 to 9. For the spherical particles of different sizes, although the particle is stabilized at different heights ($z'$-axis), they are all trapped right above the center of the transducer array. Theoretically, the radiation torque is close to zero as required by symmetry. In contrast, for the non-spherical objects, the difference of the geometrical features strongly affects the scattering properties around the particles, thus changing the radiation force and torque, thereby the motion of the particles. It can be found that both large and small ellipsoidal particles are difficult to capture (moving below the critical lines of $z' = -50$ mm). Compared with other geometric shapes, although they all have the same averaged radius $a$, the ellipsoidal particles tend to alter their orientation, resulting in the effective cross-sectional area facing the wavefront being the smallest. Hence, the radiation force $F_{\text{rad},z'}$ acting on the elliposoidal particles is relatively small and insufficient to offset the gravity. For the small cone and diamond particles, their translational and rotational motions are basically identical to the spherical particles. With the increase of particle sizes, geometric features become an indispensable factor. The geometric asymmetry with respect to the wavefront induces the additional radiation force and torque, which translate and rotate the cone and diamond particles along different routes. It should be emphasized that when the particle size parameter reaches a certain level ($ka$~1), the radiation effect remains unchanged [24]. However, the gravity is proportional to the cube of the averaged radius ($F_G \propto a^3$). Hence, with the increase of particle size, the radiation force is not enough to offset the gravity. Comparing the results given in Figs. 8 and 9, it can be found that the cone particle with an averaged radius of $a = 2$ mm can be stably trapped, while the transducer array fails to capture the cone particle when its averaged radius is increased to $a = 3$ mm. Finally, it can be seen that the large non-spherical particles are tended to be trapped at $(x', y') = (\pm 5, \pm 5)$ mm. The difference is that smaller particles prefer to stabilize at right above the center of the transducer array, that is $(x', y') = (0,0)$ mm. This can be explained by tha fact that the scattered effects due to the geometric differences become insignificant for the small objects. As a result, the motions of the small non-spherical particles are close to the motions of their spherical counterparts.

According to the predictions (Figs. 6 to 9) and the above discussions, it can be found that the geometric asymmetry is a potential degree of freedom to tune acoustophoretic processes. The simplification of external geometry to a sphere neglects the effects of asymmetry, which may lead to a considerable deviation between experiments and expectations, especially for large objects. Here, we provide an efficient and accurate method to estimate the acoustophoretic processes, considering the significance of geometric features.

# 4. Conclusions

This paper presents a theoretical framework to predict the acoustophoretic process of any axisymmetric particles, driven by acoustic radiation force and torque, above a user-customized transducer array. We start with establishing a computation coordinate system (CCS) to facilitate the derivation of the radiation force and torque, while an observation coordinate system (OCS) to visualize the motion of a particle in the perspective of the observers. In the CCS, a semi-analytical method is proposed to estimate the acoustic radiation force and torque acting exerted on an axisymmetric particle above a user-customized transducer array in the air. The derivation is based on the conformal transformation approach, which maps the irregular surface into a spherical surface so that the boundary conditions can be employed to solve the scattered wavefield. Note that the derivation is based on the premise that the symmetric axis of the particle is parallel to the $z$-axis of the CCS (seeing Fig. 3). Therefore, it fails to the scenarios that the particle rotation deviates from the $z$-axis. To break this limitation, we reconsider the rotation of the particle in the OCS as the opposite rotation of the incident driving field (or transducer array) in the CCS (seeing Figs. 1 and 2). In this case, a rotation transformation is employed to tune the incident driving field, and the corresponding radiation force and torque are obtained under the CCS, while another rotation transformation is required to transfer the data from the CCS back to the OCS.

The performance of the framework is fully evaluated by comparing the semi-analytical results with three-dimensional numerical examples. Specifically, the radiation force and torque exerted on a non-spherical particle with different geometric features (ellipsoid, cone, and diamond) and particle orientations levitated above a transducer array with varying phase distributions are thoroughly compared. It could be found that the proposed method shows superior computational accuracy, high geometric adaptivity, and good computational robustness (Fig. 5), while requiring much less computational time (~ 5 s v.s. ~ 10 mins for numerical method) than that based on the numerical method. The translational and rotational dynamics of various particles, i.e., acoustophoresis of particles, are visualized (Figs. 6 to 9), dominated by Newton's law under the viscosity and the radiation effects. The results illustrate that shape asymmetry could be an essential factor in tuning the acoustophoretic process. Although the effect of shape asymmetry is negligible for small particles, the scattering property becomes considerable, and the geometric feature plays a vital role in the dynamics of large particles. Furthermore, a potential benefit of our method is that it is able to predict the trajectory of non-spherical particles under a user-specified wavefront, which is impractical in existing numerical simulations since they are cumbersome in continuously updating the position and orientation information of particles, and completing the mesh establish and calculation process.

The proposed framework can be an effective and efficient tool to predict the motion of various irregular objects, which helps to understand the acoustophoresis of the irregular particles over a wide range of size parameters. Additionally, incorporating the phase retreval algorithms to the framework makes it possible to achieve user-specified rotational and translational manipulation of non-spherical objects.

## Appendix A: Conformal transformation and mapping coefficients

As there should be only one value of the given slice function $r_s(\theta)$ for each $\phi$, the mapping procedure for the axisymmetric body is commenced by expanding function $r_s(\theta)$ in a Fourier series relative to the polar angle, $\phi$, as

$$r_s(\theta) = a + \sum_{n=1}^{\infty}[A_n\cos(n\theta) + B_n\sin(n\theta)], \quad (A.1)$$

where $a$ is the average radius of the body, and $A_n$ and $B_n$ are the Fourier series coefficients. Note that the Fourier expansion is performed for the period of $2\pi$, while the polar angular coordinate $\theta$ is defined from 0 to $\pi$. Consequently, although the series is intentionally computed based on the periodic extension from $\pi$ to $2\pi$, the polar angle is only meaningful in the range of $[0,\pi]$. Equation (A.1) can be rewritten in terms of exponentials as

$$r_s(\theta) = a + \sum_{n=1}^{\infty}\left[R_n^* e^{n\theta i} + R_n e^{-n\theta i}\right], \quad (A.2)$$

where $R_n = \frac{1}{2}(A_n + B_n)$ and the superscript symbol $*$ means taking conjugation of the corresponding variable. It is convenient to describe the boundary of the slice using the complex system

$$r_s(\theta)e^{\theta i} = ae^{\theta i} + \sum_{n=1}^{\infty}\left[R_n^* e^{(1+n)\theta i} + R_n e^{(1-n)\theta i}\right]. \quad (A.3)$$

The real part of $r_s(\theta)e^{\theta i}$ corresponds to the projection value of $r_s(\theta)$ in the $z$-axis and imaginary part of $r_s(\theta)e^{\theta i}$ is the projection value in $f$-axis for the real $zOf$ plane in Fig. 3.

On the boundary of the slice of the irregular particle, we equating $r_s(\theta)e^{\theta i}$ in Eq. (A.3) to complex mapping function $M(u + w \cdot i)$ in Eq. (6) with $u = u_0 = 0$ yields

$$ae^{\theta i} + \sum_{n=1}^{\infty}\left[R_n^* e^{(1+n)\theta i} + R_n e^{(1-n)\theta i}\right] = c_{-1}e^{wi} + \sum_{n=0}^{\infty} c_n e^{-nwi}. \quad (A.4)$$

Since the boundary of the slice is a periodic function, the deviation of $\theta$ from $w$ can be represented as a Fourier series

$$\theta = w + \sum_{n=1}^{\infty}[E_n\cos(nw) + F_n\sin(nw)]. \quad (A.5)$$

In the above equation, the series coefficients $E_n$ and $F_n$ are unknown, while can be determined by orthogonality relationship of complex exponential functions $\int_0^{2\pi} e^{nwi} \cdot e^{-mwi} \, dw = 2\pi \delta_{n,m}$, where $\delta_{n,m}$ is the Kronecker delta function. We multiply both sides of Eq. (A.1) by $\frac{1}{2\pi} e^{-mwi}$ and integrating over $w$ from 0 to $2\pi$

$$\begin{cases} \frac{1}{2\pi} \int_2^{2\pi} e^{-mwi} \left\{ ae^{\theta i} + \sum_{n=1}^{\infty} \left[ R_n^* e^{(1+n)\theta i} + R_n e^{(1-n)\theta i} \right] \right\} dw = 0, & m > 1 \\ \frac{1}{2\pi} \int_2^{2\pi} e^{-mwi} \left\{ ae^{\theta i} + \sum_{n=1}^{\infty} \left[ R_n^* e^{(1+n)\theta i} + R_n e^{(1-n)\theta i} \right] \right\} dw = c_{-m}, & m \leq 1 \end{cases} \quad \text{(A.6)}$$

Based on Eq. (A.6), the series coefficients $E_n$ and $F_n$ can be solved using the upper equation, which are then used to obtain the mapping coefficients through the lower equation.

## Appendix B: Dirichlet and Neumann boundary conditions

Based on the mapping relationships given in Eq. (16), the position vector can be generally expressed as

$$\vec{r} = f(u,w) \cdot \cos(v) \vec{e}_x + f(u,w) \cdot \sin(v) \vec{e}_y + g(u,w) \vec{e}_z, \quad \text{(B.1)}$$

where $\vec{e}_x$, $\vec{e}_y$, and $\vec{e}_z$ are unit vectors along the corresponding coordinate axes. In the new coordinate system, the orthogonal coordinate system is desirable since it facilitates the computation of the normal particle velocity on the boundary. Orthogonality of the new coordinate system requires that the partial derivative of the position vector $\vec{r}$ in Eq. (B.1) satisfies

$$\vec{r}_u \cdot \vec{r}_w = 0; \quad \vec{r}_u \cdot \vec{r}_v = 0; \quad \vec{r}_v \cdot \vec{r}_w = 0, \quad \text{(B.2)}$$

where the subscripts mean the partial derivative of corresponding variables. Considering the mapping relationship given in Eq. (B.1), the partial derivatives of the position vector with respect to each of the variables are

$$\begin{cases} \vec{r}_u = f_u(u,w) \cdot \cos(v) \vec{e}_x + f_u(u,w) \cdot \sin(v) \vec{e}_y + g_u(u,w) \vec{e}_z \\ \vec{r}_w = f_w(u,w) \cdot \cos(v) \vec{e}_x + f_w(u,w) \cdot \sin(v) \vec{e}_y + g_w(u,w) \vec{e}_z, \\ \vec{r}_v = -f(u,w) \cdot \sin(v) \vec{e}_x + f(u,w) \cdot \cos(v) \vec{e}_y \end{cases} \quad \text{(B.3)}$$

Inserting Eq. (B.3) into Eq. (B.2), it can be proven that the new coordinate system could be orthogonal if the mapping functions satisfy

$$f_u(u,w) = g_w(u,w) \text{ or } f_w(u,w) = g_u(u,w). \quad \text{(B.4)}$$

The Dirichlet boundary condition requires that the total potential vanishes on the surface of the scatterer $\hat{\phi}_{ex}(u_0, w, v) + \hat{\phi}_{sc}(u_0, w, v) = 0$ (derived from Eqs. (18) and (19)), which gives

$$\sum_{n,m} a_{nm} J_n^m(u_0, w, v) + \sum_{n,m} s_{nm} a_{nm} H_n^m(u_0, w, v) = 0. \tag{B.5}$$

The system of equations necessary to satisfy this boundary condition is generated by multiplying both sides of this equation by a set of spherical angular eigenfunctions

$$\psi_{n'}^{m'}(w, v) = P_{n'}^{m'}(\cos(w)) \sin(w) e^{-m'v i}, \tag{B.6}$$

and integrating over the range of $w$ and $v$ [41]:

$$\int_0^\pi \int_0^{2\pi} \left[ \sum_{n,m} a_{nm} J_n^m(u_0, w, v) \right. \tag{B.7}$$

$$\left. + \sum_{n,m} s_{nm} a_{nm} H_n^m(u_0, w, v) \right] \psi_{n'}^{m'}(w, v) dv dw = 0.$$

Considering the orthogonality relationship $\int_0^{2\pi} e^{nw i} \cdot e^{-mw i} dw = 2\pi \delta_{n,m}$ and the definition of spherical harmonic function $Y_n^m(\theta, \varphi) = \sqrt{\frac{(2n+1)}{4\pi} \cdot \frac{(n-m)!}{(n+m)!}} P_n^m(\cos(\theta)) e^{m\varphi i}$ [39], the above equation becomes

$$\sum_{n=0}^N a_{nm'} \Gamma_n^{n',m'} + \sum_{n=0}^N s_{nm'} a_{nm'} \Lambda_n^{n',m'} = 0, \tag{B.8}$$

$$(n' = 0,1,\cdots,N; m' = -N,\cdots,0,\cdots,N)$$

where the structural functions $\Gamma_n^{n',m'}$ and $\Lambda_n^{n',m'}$ are

$$\begin{cases} \Gamma_n^{n',m'} = \int_0^\pi [j_n(kr(u_0, w)) \sqrt{\frac{(2n+1)}{4\pi} \cdot \frac{(n-m)!}{(n+m)!}} P_n^{m'}(\cos\theta(u_0, w)) \\ \qquad\qquad P_{n'}^{m'}(\cos(w)) \sin(w)] dw \\ \Lambda_n^{n',m'} = \int_0^\pi [h_n(kr(u_0, w)) \sqrt{\frac{(2n+1)}{4\pi} \cdot \frac{(n-m)!}{(n+m)!}} P_n^{m'}(\cos\theta(u_0, w)) \\ \qquad\qquad P_{n'}^{m'}(\cos(w)) \sin(w)] dw \end{cases}. \tag{B.9}$$

The Neumann boundary condition requires that the normal particle velocity vanishes on the scatterer surface $\vec{n} \cdot \nabla[\hat{\phi}_{ex}(u_0, w, v) + \hat{\phi}_{sc}(u_0, w, v)] = 0$, where $\vec{n}$ is the outer normal vector to the surface. It can be found that the mapping functions given in Eq. (15) satisfy the orthogonal requirements in Eq. (B.4). Consequently, the gradient of the potential field is

$$\nabla \hat{\phi}(u_0, w, v) = \frac{\partial \hat{\phi}(u_0, w, v)}{\partial u} \frac{\vec{r}_u}{r_u} + \frac{\partial \hat{\phi}(u_0, w, v)}{\partial w} \frac{\vec{r}_w}{r_w} + \frac{\partial \hat{\phi}(u_0, w, v)}{\partial v} \frac{\vec{r}_v}{r_v}, \tag{B.10}$$

where vectors $\vec{r}_u$, $\vec{r}_w$ and $\vec{r}_v$ are given in Eq. (B.3). As the scatterer surface has been defined by $u = u_0 = 0$, the outer normal vector $\vec{n}$ is parallel to $\vec{r}_u$. Hence, the Neumann boundary condition becomes

$$\frac{1}{\sqrt{f_u^2 + f_w^2}} \frac{\partial [\hat{\phi}_{\text{ex}}(u_0, w, v) + \hat{\phi}_{\text{sc}}(u_0, w, v)]}{\partial u} = 0. \tag{B.11}$$

Inserting Eqs. (18) and (19) into the above equation, multiplying both sides by the spherical angular eigenfunctions and considering the orthogonality relationship $\int_0^{2\pi} e^{nwi} \cdot e^{-mwi} dw = 2\pi \delta_{n,m}$, we finally yield

$$\sum_{n=0}^{N} a_{nm'} \Gamma_{n,u}^{n',m'} + \sum_{n=0}^{N} s_{nm'} a_{nm'} \Lambda_{n,u}^{n',m'} = 0, \tag{B.12}$$
$$(n' = 0,1,\cdots,N; m' = -N,\cdots,0,\cdots,N)$$

where $\Gamma_{n,u}^{n',m'}$ and $\Lambda_{n,u}^{n',m'}$ are the partial derivative of the new radial coordinate of structural functions $\Gamma_n^{n',m'}$ and $\Lambda_n^{n',m'}$ given in Eq. (B.9):

$$\begin{cases} \Gamma_{n,u}^{n',m'} = \left. \frac{\partial \Gamma_n^{n',m'}}{\partial u} \right|_{u=u_0} \\ \Lambda_{n,u}^{n',m'} = \left. \frac{\partial \Lambda_n^{n',m'}}{\partial u} \right|_{u=u_0} \end{cases}. \tag{B.13}$$

## Appendix C: Solution of the system of equations

Based on Eq. (21) (or Eq. (22) that follows a similar process as given below), for each combination of $(n', m')$, we can obtain an additional equation to close the system. There are totally $(N + 1) \times (2N + 1)$ additional equations and $(N + 1) \times (2N + 1)$ unknown scattering coefficients $s_{n'm'}$. For a fixed index of $m'$, the change of index of $n' = 0,1,\cdots,N$ is able to provide $N + 1$ additional equations as

$$\begin{bmatrix} a_{0m'}\Lambda_0^{0,m'} & a_{1m'}\Lambda_1^{0,m'} & \cdots & a_{Nm'}\Lambda_N^{0,m'} \\ a_{0m'}\Lambda_0^{1,m'} & a_{1m'}\Lambda_1^{1,m'} & \cdots & a_{Nm'}\Lambda_N^{1,m'} \\ \vdots & \vdots & \ddots & \vdots \\ a_{0m'}\Lambda_0^{N,m'} & a_{1m'}\Lambda_1^{N,m'} & \cdots & a_{Nm'}\Lambda_N^{N,m'} \end{bmatrix} \cdot \begin{bmatrix} s_{0m'} \\ s_{1m'} \\ \vdots \\ s_{Nm'} \end{bmatrix} = \begin{bmatrix} A^{0,m'} \\ A^{1,m'} \\ \vdots \\ A^{N,m'} \end{bmatrix}, \quad (C.1)$$

where abbreviation $A^{n',m'} = \sum_{n=0}^{N} a_{nm'}\Gamma_n^{n',m'}$. Solving the above linear equations can get $N+1$ scattering coefficients $s_{n'm'}$ $(n' = 0,1,\cdots,N)$. The change of index of $m'$ from $-N$ to $N$ gives a total of $2N+1$ linear systems, and therefore all the unknown scattering coefficients $s_{n'm'}$ $(n' = 0,1,\cdots,N; m' = -N,\cdots,0,\cdots,N)$ can be determined by solving $2N+1$ linear systems, corresponding to different indexes of $m'$.

## Appendix D: Numerical evaluation of radiation force and torque

The acoustic radiation force and torque on an object due to scattering phenomena was obtained as a surface integration of the object [1][2][3]

$$\vec{F}_{rad} = \int_R \langle L \rangle d\vec{A}_R - \rho_0 \int_R d\vec{A}_R \cdot \langle \vec{u}\vec{u} \rangle, \quad (D.1)$$

and

$$\vec{T}_{rad} = -\rho_0 \int_R \langle (d\vec{A}_R \cdot \vec{u}) \cdot (\vec{r} \times \vec{u}) \rangle, \quad (D.2)$$

where the angle bracket $\langle \cdot \rangle$ denotes the time average of the variable therein. $L$ is the acoustic Lagrange density defined as $L = \frac{1}{2}\rho_0 \vec{u} \cdot \vec{u} - \frac{1}{2\rho_0 c_s^2}p^2$, where $\rho_0 \vec{u} \cdot \vec{u}$ is the flux of momentum density. The spherical surface R surrounding the scattering particle is the same as defined in Eq. (6), and the direction of the integration element $d\vec{A}_R$ is along the outer normal of the surface.

Here, the outer normal vector of integrating surface R can be expressed as $d\vec{A}_R = \vec{e}_R dA_R$, where $\vec{e}_R = \left(\frac{x}{a_R}, \frac{y}{a_R}, \frac{z}{a_R}\right)$ defined as the unit outer normal vector of spherical surface R with a radius of $a_R = \sqrt{x^2+y^2+z^2}$. The point position on the integrating surface is denoted as $(x,y,z)$ under the Cartesian coordinate system. Inserting $d\vec{A}_R = \vec{e}_R dA_R$ into Eqs. (D.1) and (D.2), using tensor relation $\vec{e}_R \cdot (\vec{u}\vec{u}) = (\vec{e}_R \cdot \vec{u})\vec{u}$, we arrive at

$$\begin{cases} \vec{F}_{\mathrm{rad}} = \int_{\mathrm{R}} \langle \frac{\rho_0}{2}\vec{u}\cdot\vec{u} - \frac{1}{2\rho_0 c_s^2} p^2 \rangle \vec{e}_{\mathrm{R}} \mathrm{d}A_{\mathrm{R}} - \rho_0 \int_{\mathrm{R}} \langle (\vec{e}_{\mathrm{R}}\cdot\vec{u})\vec{u}\rangle \mathrm{d}A_{\mathrm{R}} \\ \vec{T}_{\mathrm{rad}} = -\rho_0 \int_{\mathrm{R}} \langle (\vec{e}_{\mathrm{R}}\cdot\vec{u})\cdot(\vec{r}\times\vec{u})\rangle \mathrm{d}A_{\mathrm{R}} \end{cases} \quad (\mathrm{D}.3)$$

Using the relationship $\langle XY \rangle = \frac{1}{2}\mathrm{Re}(\hat{X}\hat{Y}^*)$, the radiation force and torque are rearranged along corresponding coordinate axes under the Cartesian coordinate system as

$$F_{\mathrm{rad},x} = \vec{F}\cdot\vec{e}_x = \int_{\mathrm{R}} \frac{1}{4}\frac{x}{a_{\mathrm{R}}}\left[\rho_0 \mathrm{Re}(\hat{\vec{u}}\cdot\hat{\vec{u}}^*) - \frac{1}{\rho_0 c_s^2}\mathrm{Re}(\hat{p}\cdot\hat{p}^*)\right]\mathrm{d}A_{\mathrm{R}} \quad (\mathrm{D}.4)$$
$$- \frac{\rho_0}{2}\int_{\mathrm{R}}\left[\frac{x}{a_{\mathrm{R}}}\mathrm{Re}(\hat{u}_x\cdot\hat{u}_x^*) + \frac{y}{a_{\mathrm{R}}}\mathrm{Re}(\hat{u}_x\cdot\hat{u}_y^*) + \frac{z}{a_{\mathrm{R}}}\mathrm{Re}(\hat{u}_x\cdot\hat{u}_z^*)\right]\mathrm{d}A_{\mathrm{R}},$$
$$F_{\mathrm{rad},y} = \vec{F}\cdot\vec{e}_y = \int_{\mathrm{R}} \frac{1}{4}\frac{y}{a_{\mathrm{R}}}\left[\rho_0 \mathrm{Re}(\hat{\vec{u}}\cdot\hat{\vec{u}}^*) - \frac{1}{\rho_0 c_s^2}\mathrm{Re}(\hat{p}\cdot\hat{p}^*)\right]\mathrm{d}A_{\mathrm{R}}$$
$$- \frac{\rho_0}{2}\int_{\mathrm{R}}\left[\frac{x}{a_{\mathrm{R}}}\mathrm{Re}(\hat{u}_y\cdot\hat{u}_x^*) + \frac{y}{a_{\mathrm{R}}}\mathrm{Re}(\hat{u}_y\cdot\hat{u}_y^*) + \frac{z}{a_{\mathrm{R}}}\mathrm{Re}(\hat{u}_y\cdot\hat{u}_z^*)\right]\mathrm{d}A_{\mathrm{R}},$$
$$F_{\mathrm{rad},z} = \vec{F}\cdot\vec{e}_z = \int_{\mathrm{R}} \frac{1}{4}\frac{z}{a_{\mathrm{R}}}\left[\rho_0 \mathrm{Re}(\hat{\vec{u}}\cdot\hat{\vec{u}}^*) - \frac{1}{\rho_0 c_s^2}\mathrm{Re}(\hat{p}\cdot\hat{p}^*)\right]\mathrm{d}A_{\mathrm{R}}$$
$$- \frac{\rho_0}{2}\int_{\mathrm{R}}\left[\frac{x}{a_{\mathrm{R}}}\mathrm{Re}(\hat{u}_z\cdot\hat{u}_x^*) + \frac{y}{a_{\mathrm{R}}}\mathrm{Re}(\hat{u}_z\cdot\hat{u}_y^*) + \frac{z}{a_{\mathrm{R}}}\mathrm{Re}(\hat{u}_z\cdot\hat{u}_z^*)\right]\mathrm{d}A_{\mathrm{R}},$$

and

$$T_{\mathrm{rad},x} = \vec{T}\cdot\vec{e}_x = -\frac{\rho_0}{2}\int_{\mathrm{R}} \frac{xy}{a_{\mathrm{R}}}\mathrm{Re}(\hat{u}_x\cdot\hat{u}_z^*) + \frac{y^2-z^2}{a_{\mathrm{R}}}\mathrm{Re}(\hat{u}_y\cdot\hat{u}_z^*) \quad (\mathrm{D}.5)$$
$$+ \frac{yz}{a_{\mathrm{R}}}\mathrm{Re}(\hat{u}_z\cdot\hat{u}_z^*) - \frac{x^2}{a_{\mathrm{R}}}\mathrm{Re}(\hat{u}_x\cdot\hat{u}_y^*) - \frac{yz}{a_{\mathrm{R}}}\mathrm{Re}(\hat{u}_y\cdot\hat{u}_y^*)\mathrm{d}A_{\mathrm{R}},$$

$$T_{\mathrm{rad},y} = \vec{T}\cdot\vec{e}_y = -\frac{\rho_0}{2}\int_{\mathrm{R}} \frac{xz}{a_{\mathrm{R}}}\mathrm{Re}(\hat{u}_x\cdot\hat{u}_x^*) + \frac{yz}{a_{\mathrm{R}}}\mathrm{Re}(\hat{u}_x\cdot\hat{u}_y^*)$$
$$+ \frac{z^2-x^2}{a_{\mathrm{R}}}\mathrm{Re}(\hat{u}_x\cdot\hat{u}_z^*) - \frac{xy}{a_{\mathrm{R}}}\mathrm{Re}(\hat{u}_y\cdot\hat{u}_z^*) - \frac{xz}{a_{\mathrm{R}}}\mathrm{Re}(\hat{u}_z\cdot\hat{u}_z^*)\mathrm{d}A_{\mathrm{R}},$$

$$T_{\mathrm{rad},z} = \vec{T}\cdot\vec{e}_z = -\frac{\rho_0}{2}\int_{\mathrm{R}} \frac{x^2-y^2}{a_{\mathrm{R}}}\mathrm{Re}(\hat{u}_x\cdot\hat{u}_y^*) + \frac{xy}{a_{\mathrm{R}}}\mathrm{Re}(\hat{u}_y\cdot\hat{u}_y^*)$$
$$+ \frac{xz}{a_{\mathrm{R}}}\mathrm{Re}(\hat{u}_y\cdot\hat{u}_z^*) - \frac{xy}{a_{\mathrm{R}}}\mathrm{Re}(\hat{u}_x\cdot\hat{u}_x^*) - \frac{yz}{a_{\mathrm{R}}}\mathrm{Re}(\hat{u}_x\cdot\hat{u}_z^*)\mathrm{d}A_{\mathrm{R}}.$$

Here $\hat{p}$ and $\hat{\vec{u}} = (\hat{u}_x, \hat{u}_y, \hat{u}_z)$ are the complex amplitudes of acoustic pressure and particle velocity, respectively. $\vec{e}_x$, $\vec{e}_y$, and $\vec{e}_z$ are unit vectors along the corresponding axes.